\documentclass[12pt,a4paper,final]{article}
\usepackage[T2A]{fontenc}
\usepackage{indentfirst}
\usepackage{epsfig}
\usepackage{rotating}
\usepackage[english]{babel}
\usepackage{wrapfig}
\begin{document}

\title{ On the possibility to use ATLAS and CMS detectors for neutrino physics. }
\author{A. Guskov\\[0.3cm]
{\normalsize\it Joint Institute for Nuclear Research},\\
 {\normalsize\it 141980, Dubna, Russia}, \\
  {\normalsize\it avg@nusun.jinr.ru} }
\date{\empty}
\maketitle

\vskip 5mm
\begin{abstract} 
 Energetic primary cosmic rays entering the Earth's atmosphere generate  flux of secondary particles including neutrinos. 
Muon neutrinos passed through the Earth and  produced muons via the charged current reaction (CC) can be registered by
 experimental setups intended for the measurements with colliding beams.
 Due to large geometrical size and advanced muon detecting system such detectors as ATLAS and CMS on LHC have chance to contribute
 also into the neutrino physics. The estimation of possible rates of up-going muons produced by neutrinos  is given.
\end{abstract} 

The ATLAS \cite{ATLAS} and the CMS \cite{CMS} detectors are the general purpose detectors on the Large Hadron Collider. They are intended to study particle interaction 
at high energies. But the universality and $4\pi$ geometry of these detectors provide possibility to measure the flux of the atmospheric
muon neutrinos via the registration of up-going muons, produced in the charged current neutrino interactions in surrounding rock, by the muon system of the detectors.

\begin{figure}[h]
\begin{center}
\includegraphics[width=30pc]{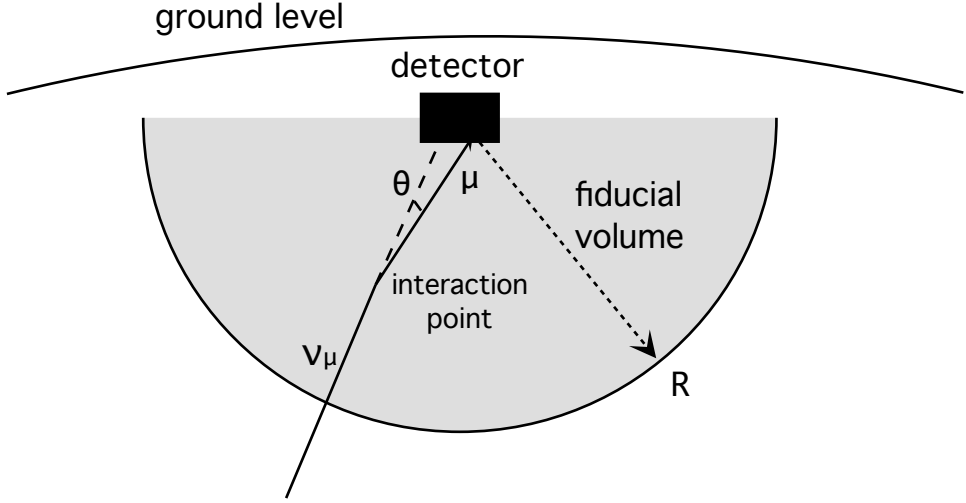}
\caption{\label{pic:0}Use of the detectors like ATLAS and CMS as neutrino telescopes.}
\end{center}
\end{figure}

The event rate of up-going muons can be expressed as:
\begin{equation}
\frac{dN}{dt} = \int w_{int} \frac{d^4\Phi}{dE_{\nu}dSd\Omega dt} dE_{\nu}dSd\Omega,
\end{equation}
where $w_{int}$ is the probability of neutrino CC interaction in fiducial volume of the detector. The fiducial volume (see Fig. \ref{pic:0}) of the detector can be characterized by the aperture
$\Omega_0(E_{\nu})$ and the radius $R(E_{\nu})$. In the simplest case the aperture can be assumed to be independent on the neutrino energy  and equal to $2\pi$ (lower hemisphere). The radius $R(E_{\nu})$ can be defined as the maximal distance which can be passed by produced muon to be detected:
\begin{equation}
R(E_{\nu})= r(E_{\mu})-r(E_{min}),
\end{equation}
where $r$ is average muon range and $E_{min}$ is the threshold of muon registration. The muon energy loss may be expressed as
\begin{equation}
-dE/dX = a+b E_{\mu},
\end{equation}
where $a$ is the ionization loss and $b$ is the energy loss by the radiative processes. So the average muon range is:
\begin{equation}
r(E_{\mu})=\frac{1}{b}ln(1+\frac{bE_{\mu}}{a})
\end{equation}

The probability of muon production within the fiducial volume is:
\begin{equation}
w_{int} = \sigma n_{N} R(E_{\nu}),
\end{equation}
where $\sigma$ is the CC cross section and $n_{N}$ is the concentration of nucleons.

\begin{figure}[h]
\begin{minipage}{14pc}
\includegraphics[width=14pc]{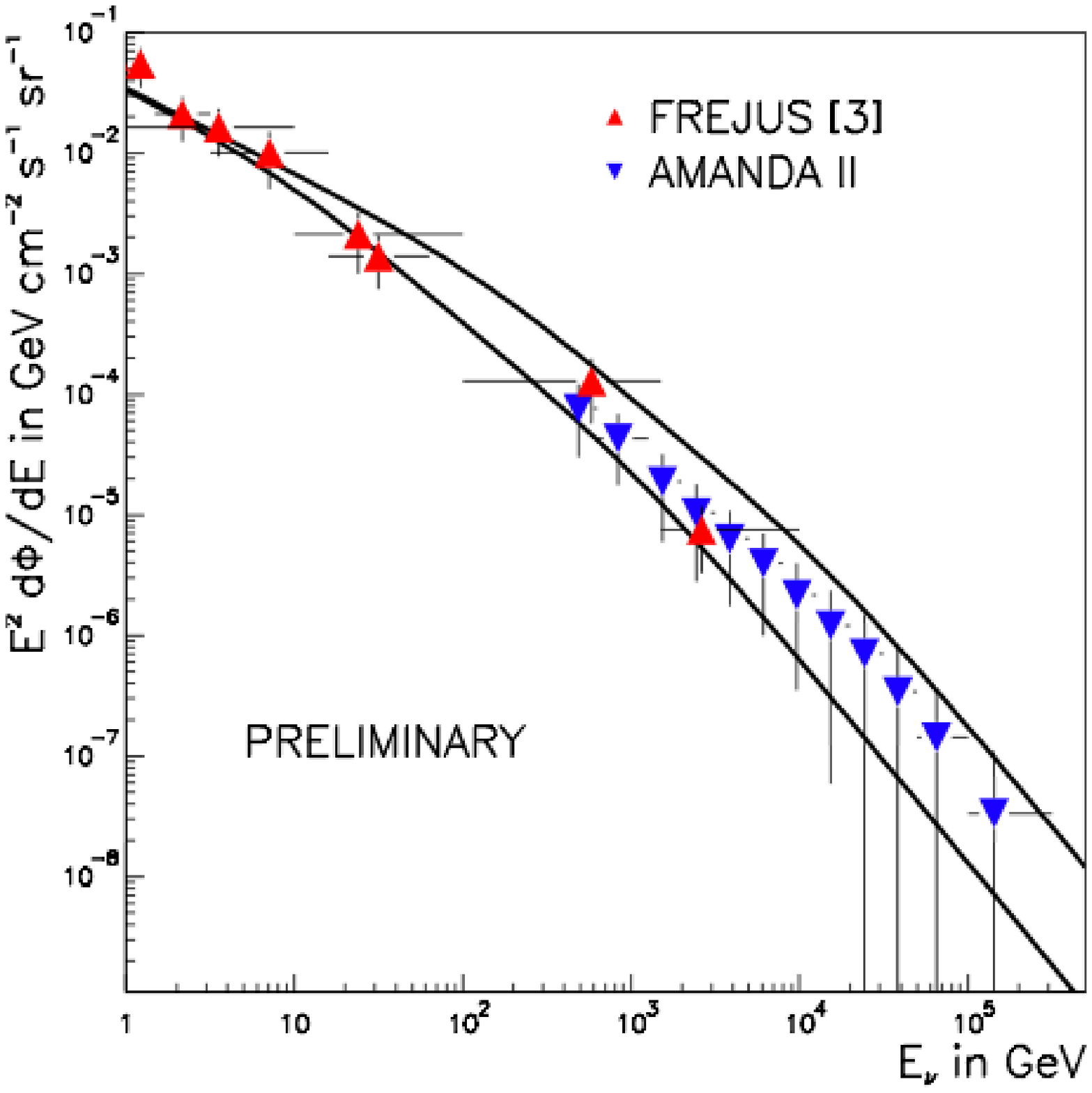}
\caption{\label{pic1}The measured flux of the atmospheric $\nu_{\mu}$ \cite{AMANDA}. }
\end{minipage} \hspace{2pc}%
\begin{minipage}{22pc}
\includegraphics[width=22pc]{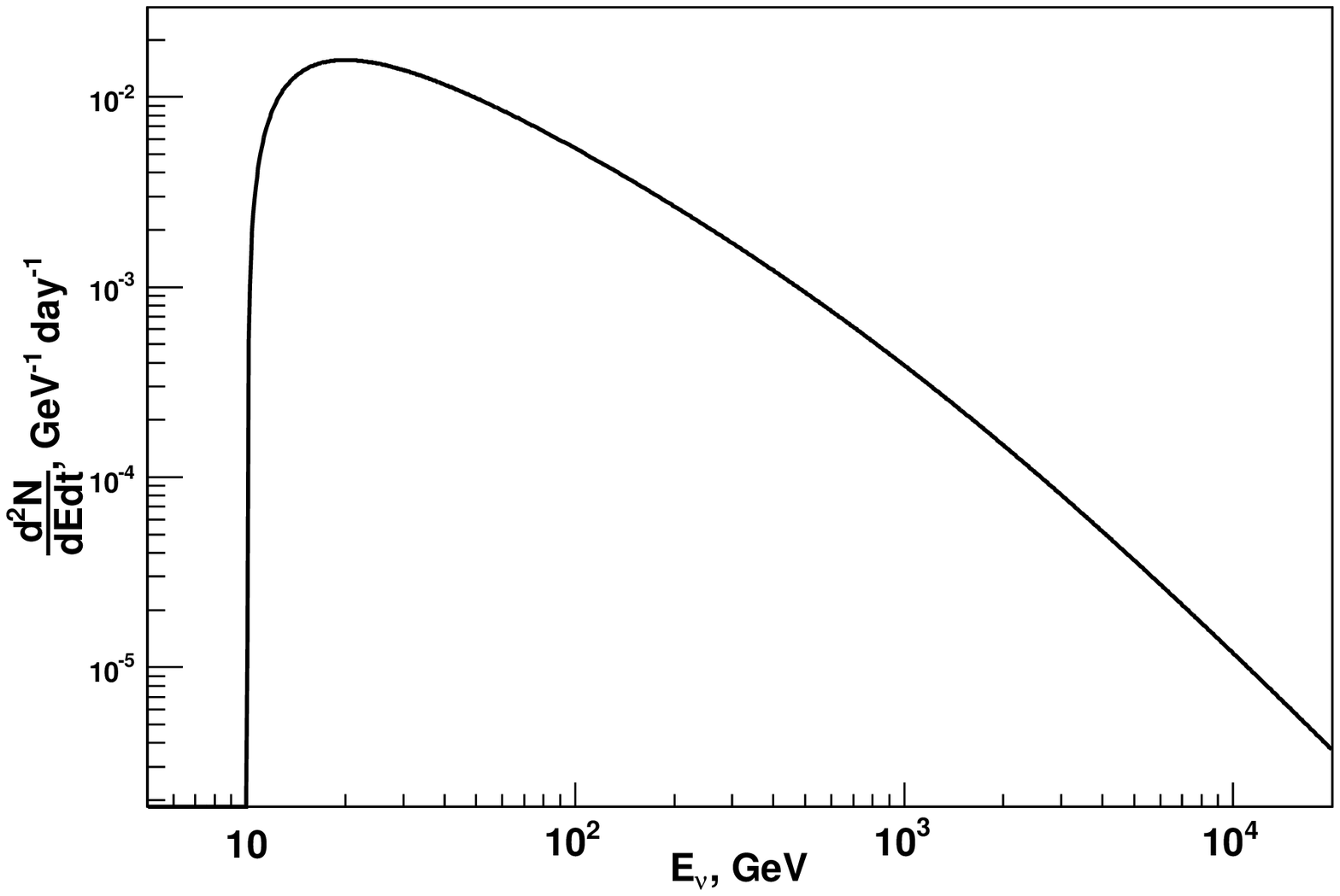}
\caption{\label{pic:2} Spectral sensitivity of ATLAS detector.}
\end{minipage} \hspace{1pc}
\end{figure}

The flux of the muon neutrinos  in the energy above 10 GeV (see  \cite{AMANDA} and Fig. \ref{pic1})can be parameterized by the expression:

\begin{equation}
\frac{d\Phi}{dE_{\nu}dSd\Omega dt} = 0.1\times E_{\nu}^{-3} ~GeV^{-3}cm^{-2}sr^{-1}s^{-1}.
\end{equation}
The averaged CC cross section for $\mu_{mu}$ and $\bar{\nu}_{\mu}$ per nucleon is:
\begin{equation}
\sigma=\frac{\sigma_{\nu_{\mu}}+\sigma_{\nu_{\bar{\mu}}}}{2}=0.5\times10^{-38}\times E_{\nu}[GeV]~ cm^{2} ~ \cite{PDG}.
\end{equation}

For numerical estimation one can assume that the density of surrounding rock is $2.65~ g \times cm^{-3}$, the concentration of the nucleons $n_{N}$
to be about $1.6\times10^{24} ~cm^{-3}$, $a=6.5\times 10^{-3} ~GeV/cm$ , $b=10^{-5} ~ cm^{-1}$ and $E_{\mu}\approx E_{\nu}$. One also assumes that the effective square $S_{eff}$ of the detectoris
$10^{7}~cm^2$ and the threshold of muon registration $E_{min}$=10 GeV. Such parameters look like to be realistic for ATLAS detector.
So, for 100\% efficiency of trigger and muon track reconstruction the event rate of up-goung muons produced by neutrinos is: 
\begin{equation}
\frac{dN}{dt}=400 \int_{10 GeV}\frac{1}{E_{\nu}^{2}}(ln(1-0.0015 E_{\nu})-0.015)dE_{\nu}=2.5 ~day^{-1}.
\end{equation}

 The energy distribution of atmospheric neutrinos which can be detected by ATLAS via CC reaction is shown in Fig \ref{pic:2}. The fiducial volume for $E_{\nu}=100 ~GeV$ is $7\times10^6~ m^{3}$.

The event rate for CMS detector is less due to smaller geometrical size of the detector but also about 1 event per day. The estimated rates of neutrino events about 1 event per day are comparable with the event rate in the AMANDA2 experiment \cite{AMANDA2} ($\approx 2~ day^{-1}$) and in the  Super-Kamiokande experiment ($\approx 1.5~ day^{-1}$) \cite{SK}. Therefore ATLAS and CMS detectors can be successfully used for such tasks of neutrino physics like study of atmospheric neutrinos oscillation and registration neutrinos from astrophysical sources. Unquestionable advantage of the detectors like ATLAS and CMS is the possibility to distinguish between events induced by neutrino and antineutrino. So the differential fluxes of $\nu_{\mu}$ and $\bar{\nu}_{\mu}$ can be measured independently.

The main shortcoming of LHC detectors for neutrino physics is that they are places at low depth under the ground level. 300 m of water equivalent is not enough to protect the detector against the muons coming from upper hemisphere and scattered back. (This value should be compared with 1-2 km for AMANDA, 1.3-2.3 km for IceCube \cite{IC} and 2.7 km for Super-Kamiokande.) But one can hope that contributions from background muons from upper hemisphere and from the muons induced by neutrino interactions have different energy and angular distributions and can be separated statistically in offline analysis.

The precision of measurement of neutrino direction is limited by the uncertainty of the angle between neutrino and muon going directions which is about $3.5^{0}$ for $E_{\mu}=100~ GeV$ while the upper limit of  uncertainty related with the multiple scattering in rock is $2^{0}$.\\

{\bf Acknowledgments.} 
I wish to thank G. Chelkov, P. Nevsky and C. Spiering for their interest to this work and comprehensive discussion.

\end{document}